\documentclass[12pt]{article}
\usepackage{amsmath,bbm,amssymb}
\usepackage{amsthm}
\usepackage{comment}
\usepackage{graphicx, graphics}
\usepackage{textcomp}
\usepackage{amsfonts}

\usepackage{psfrag,pst-node,subfigure,rotating, amsmath, bbm, amsthm, amssymb, amsthm, setspace, picture, epsfig, amsfonts, upgreek}

\usepackage{multirow}

\usepackage{arydshln}

\newtheorem{assumption}{Assumption}
\newtheorem{corollary}{Corollary}
\newtheorem{theorem}{Theorem}
\newtheorem{lemma}{Lemma}

\newtheorem{proposition}{Proposition}

\def\v{\mathbf}\def\m{\mathbf}\def\vg{\boldsymbol}\def\s{\mathcal}
\def\P{{\mathrm{P}}}\def\E{{\mathrm{E}}}

\begin{document}
\begin{center}
    {\Large\bf Better subset regression}
\\[2mm] {\large Shifeng Xiong}
\\ Academy of Mathematics and Systems Science \\Chinese Academy of Sciences, Beijing 100190\\xiong@amss.ac.cn
\end{center}

\vspace{1cm} \noindent{\bf Abstract}\quad To find efficient screening methods for high dimensional linear regression models, this paper studies the relationship between
model fitting and screening performance. Under a sparsity assumption, we show that a subset that includes the true submodel always yields smaller residual sum of squares
(i.e., has better model fitting) than all that do not in a general asymptotic setting. This indicates that, for screening important variables, we could follow a ``better
fitting, better screening" rule, i.e., pick a ``better" subset that has better model fitting. To seek such a better subset, we consider the optimization problem
associated with best subset regression. An EM algorithm, called orthogonalizing subset screening, and its accelerating version are proposed for searching for the best
subset. Although the two algorithms cannot guarantee that a subset they yield is the best, their monotonicity property makes the subset have better model fitting than
initial subsets generated by popular screening methods, and thus the subset can have better screening performance asymptotically. Simulation results show that our
methods are very competitive in high dimensional variable screening even for finite sample sizes.

\vspace{1cm} \noindent{\bf KEY WORDS:} Best subset regression; Combinatorial optimization; Dimensionality reduction; EM algorithm; Orthogonal design; Sure screening
property; Variable selection.

\newpage

\section{Introduction}\label{sec:intro}

Regression problems with large numbers of candidate predictive variables occur in a wide variety of scientific fields, and then become increasingly important
in statistical research. Suppose that there are $p$ predictive variables $X_1,\ldots,X_p$. Consider a linear regression model
\begin{equation}\label{lm}\v{y}=\m{X}\vg{\beta}+\vg{\varepsilon},\end{equation}
where $\m{X}=(x_{ij})$ is the $n\times p$ regression matrix, $\v{y}=(y_1,\ldots,y_n)'\in{\mathbb{R}}^n$ is the response vector, $\vg{\beta}=(\beta_1,\ldots,\beta_p)'$ is
the vector of regression coefficients corresponding to $X_1,\ldots,X_p$ and $\vg{\varepsilon}=(\varepsilon_1,\ldots,\varepsilon_n)'$ is a vector of independent and
identically distributed random errors with zero mean and finite variance $\sigma^2$. Without loss of generality, assume that $\m{X}$ is standardized with
$\sum_{i=1}^nx_{ij}=0$ and $\sum_{i=1}^nx_{ij_1}^2=\sum_{i=1}^nx_{ij_2}^2$ for any $j,j_1,j_2\in\{1,\ldots,p\}$ and that $\v{y}$ is centred with $\sum_{i=1}^ny_i=0$.
Throughout this paper, we denote the full model $\{1,\ldots,p\}$ by $\mathbb{Z}_p$. For $\s{A}\subset \mathbb{Z}_p$, $\m{X}_{\s{A}}$ denotes the submatrix of $\m{X}$
corresponding to $\s{A}$. For $\vg{\theta}\in\mathbb{R}^p$, $\vg{\theta}_{\s{A}}$ denotes  the subvector of $\vg{\theta}$ corresponding to $\s{A}$. For a vector $\v{x}$,
$\|\v{x}\|$ denotes its Euclidean norm. For a set $\s{S}$, $|\s{S}|$ denotes its cardinality.

With a large number of variables in (\ref{lm}), model interpretability becomes important in statistical applications. We often would like to eliminate the least
important variables for determining a smaller subset that exhibit the strongest effects. An increasing number of papers have studied on (\ref{lm}) with the sparsity
assumption that only a small number of variables among $X_1,\ldots,X_p$ contribute to the response. If the underlying model is actually sparse, the prediction accuracy
can be improved by effectively identifying the subset of important variables. When $p$ is much larger than $n$, Fan and Lv (2008) proposed a two-stage procedure for
estimating the sparse parameter $\vg{\beta}$. In the first stage, a screening approach is applied to pick $M$ variables, where $M<n$ is a specified number. In the second
stage, the coefficients in the screened $M$-dimensional submodel can be estimated by well-developed regression techniques for situations where the variables are fewer
than the observations. To guarantee the effectiveness of this procedure, the screening approach used in the first stage should possess the sure screening property, i.e.,
it retains all important variables in the model asymptotically (Fan and Lv 2008). A number of screening approaches have been studied in the literature; see Fan and Lv
(2008), Hall and Miller (2009), Fan, Samworth, and Wu (2009), Wang (2009), Fan and Song (2010), and Li, Peng, Zhang, and Zhu (2012) among others.

This paper aims to provide some new viewpoints on variable screening when $p$ is much larger than $n$. We first investigate the relationship between model fitting and
screening performance. Here model fitting of a submodel is described by the magnitude of the (residual) sum of squares it yields. Small sum of squares corresponds to
good model fitting. Consider the following question: if a submodel has better model fitting, can we say that the submodel is more likely to include all important
variables? Interestingly, the answer is ``yes" in a general asymptotic setting. The answer provides us a rule to screen variables, i.e., we should pick a submodel with
good model fitting. We call this rule ``better fitting, better screening". To make it clear, let $\s{A}_0$ denote the true submodel $\{j\in\mathbb{Z}_p:\ \beta_j\neq0\}$
with $d=|\s{A}_0|$. With a specified $M\geqslant d$, let $\mathfrak{A}_0$ and $\mathfrak{A}_1$ denote the sets $\{\s{A}\subset\mathbb{Z}_p:\ |\s{A}|=M,\
\s{A}_0\subset\s{A}\}$ and $\{\s{A}\subset\mathbb{Z}_p:\ |\s{A}|=M,\ \s{A}_0\setminus\s{A}\neq\emptyset\}$, respectively. The ``better fitting, better screening" rule
tells us that the sum of squares from a submodel $\s{A}\in\mathfrak{A}_0$ is asymptotically smaller than that from any $\s{A}\in\mathfrak{A}_1$ under regularity
conditions. In other words, a submodel of size $M$ can include $\s{A}_0$ asymptotically if it is better than $|\mathfrak{A}_1|$ other submodels of size $M$ in the sense
of model fitting. Therefore, for two subsets with the same size $M$, the better one is more likely to include $\s{A}_0$ asymptotically.

In practice, how do we find one of these better subsets? Let us consider the following optimization problem
\begin{equation}\min_{\vg{\beta}}\|\v{y}-\m{X}\vg{\beta}\|^2\quad\text{subject to}\ \|\vg{\beta}\|_0\leqslant M,\label{bss}\end{equation}where $\|\cdot\|_0$ denotes the
$\ell_0$ norm that refers to the number of nonzero components. This problem is equivalent to a combinatorial optimization problem
\begin{equation}\min_{\s{A}\subset\mathbb{Z}_p}\|\v{y}-\m{X}_{\s{A}}\hat{\vg{\beta}}_{\s{A}}\|^2\quad\text{subject to}\ |\s{A}|=M,\label{bss2}\end{equation} where
$\hat{\vg{\beta}}_{\s{A}}$ is the least squares estimator under the submodel $\s{A}$. Therefore the solution to (\ref{bss}) yields the best subset of size $M$. We can
use efficient algorithms for solving (\ref{bss}) to approximate the best subset. Even though such algorithms seldom reach the (global) solution, local solutions with
small sums of squares, which have good screening performance as well, can be obtained.

For small $p$, an exhaustive search over all possible subsets can be used to solve (\ref{bss}). A branch-and-bound strategy has been developed to reduce the number of
subsets being searched; see Beale, Kendall and Mann (1967), Hocking and Leslie (1967), LaMotte and Hocking (1970), Furnival and Wilson (1974) and Narendra and Fukunaga
(1977). Some later improvements can be found in Gatu and Kontoghiorghes (2006) and references therein. When $p$ is moderate or large, such subset searches are
infeasible. Some simplified procedures like forward stepwise selection (abbreviated as FS) can be used to give sub-optimal solutions; see e.g. Miller (2002). Note that
the solution to (\ref{bss}) has a closed form when the regression matrix $\m{X}$ is (column) orthogonal. In Section \ref{sec:os} we provide an EM algorithm to solve
(\ref{bss}). The basic idea behind this algorithm is active orthogonalization (Xiong, Dai and Qian, 2011), which embeds the original problem into a missing data problem
with a larger orthogonal regression matrix. We call this algorithm orthogonalizing subset screening (OSS). As an EM algorithm, OSS possesses the monotonicity property,
i.e., the sum of squares is not increased after an iteration. Therefore, for any sparse estimator, OSS can be used to improve its fitting by putting it as an initial
point. By the ``better fitting, better screening" rule, the screening performance can be improved as well. An accelerating algorithm, called fast orthogonalizing subset
screening, is also provided. Simulations and a real example are presented to evaluate our methods. All proofs in this paper are presented in the Appendix.

\section{The ``better fitting, better screening" rule}\label{sec:bbp}

When the underlying model (\ref{lm}) is actually sparse, it is desirable to screen $M$ variables that include all important variables. In this section we discuss the
``better fitting, better screening" rule for this purpose.

We denote any generalized inverse of a matrix $\m{A}$ by $\m{A}^-$, i.e., $\m{A}^-$ satisfies $\m{A}\m{A}^-\m{A}=\m{A}$. Note that for a submodel $\s{A}$, the least
squares estimator $(\m{X}_{\s{A}}'\m{X}_{\s{A}})^-\m{X}_{\s{A}}'\v{y}$ is not unique if $\m{X}_{\s{A}}$ is not of full (column) rank. We write
$\hat{\vg{\theta}}=(\m{X}_{\s{A}}'\m{X}_{\s{A}})^-\m{X}_{\s{A}}'\v{y}$ for meaning that $\hat{\vg{\theta}}$ belongs to the set
$\{(\m{X}_{\s{A}}'\m{X}_{\s{A}})^-\m{X}_{\s{A}}'\v{y}:\
(\m{X}_{\s{A}}'\m{X}_{\s{A}})(\m{X}_{\s{A}}'\m{X}_{\s{A}})^-(\m{X}_{\s{A}}'\m{X}_{\s{A}})=(\m{X}_{\s{A}}'\m{X}_{\s{A}})\}$. For $\s{A}\subset\mathbb{Z}_p$, let
$\hat{\vg{\beta}}^{\s{A}}$ denote the vector with $\hat{\vg{\beta}}_{\s{A}}^{\s{A}}=(\m{X}_{\s{A}}'\m{X}_{\s{A}})^{-}\m{X}_{\s{A}}'\v{y}$ and
$\hat{\vg{\beta}}_{\mathbb{Z}_p\setminus\s{A}}^{\s{A}}=\v{0}$. In this section we let $\vg{\beta}$ denote the true parameter in model (\ref{lm}). We denote by
$\lambda_{\max}(\cdot)$ and $\lambda_{\min}(\cdot)$ the largest and smallest eigenvalues of a matrix respectively. The notation $\s{A}_0$, $d$, $\mathfrak{A}_0$, and
$\mathfrak{A}_1$ are defined the same as in Section \ref{sec:intro}.

\begin{assumption}The random error $\vg{\varepsilon}$ in (\ref{lm}) follows a normal distribution $N(\v{0}, \sigma^2\m{I})$, where $\m{I}$ denotes
the identity matrix.\label{as:norm}\end{assumption}

\begin{assumption}There exists a constant $C>0$ such that $\sum_{i=1}^nx_{ij}^2/n\leqslant C$ for any $j\in\s{A}_0$.\label{as:stand}\end{assumption}
In practice, the regression matrix $\m{X}$ is usually standardized with $\sum_{i=1}^nx_{ij}^2/n=1$ for any $j\in \mathbb{Z}_p$, and then Assumption \ref{as:stand} holds.

Let $\beta_{\min}$ denote the component of $\vg{\beta}_{\s{A}_0}$ that has the smallest absolute value. To make $\s{A}_0$ well-defined, we require that $\m{X}_{\s{A}_0}$
is of full column rank and that any column in $\m{X}_{\s{A}_0}$ cannot be a linear combination of columns in $\m{X}_{\s{A}}$ for any $\s{A}\in \mathfrak{A}_1$. The two
requirements are equivalent to
$$\delta_n:=\min_{\s{A}\in \mathfrak{A}_1}\left[\frac{1}{n}\lambda_{\min}(\m{X}_{\s{A}_0\setminus\s{A}}' \m{H}_{\s{A}}\m{X}_{\s{A}_0\setminus\s{A}})\right]>0,$$where
$\m{H}_{\s{A}}=\m{I}-\m{X}_{\s{A}}(\m{X}_{\s{A}}'\m{X}_{\s{A}})^{-}\m{X}_{\s{A}}'$ is the projection matrix on the subspace $\{\v{x}\in\mathbb{R}^n:\
\m{X}_{\s{A}}'\v{x}=\v{0}\}$. The following assumption requires that $\delta_n$ (with the weakest signal $|\beta_{\min}|$) cannot converge to zero too fast. This paper
focuses on deterministic regression matrices, which makes our results applicable to designed covariates such as supersaturated designs (Wu 1993; Lin 1993). For random
design cases, it can be proved that $\delta_n$ is larger than a positive constant with high probability if all rows of $\m{X}$ are independent and identically
distributed from a non-degenerate $p$-dimensional normal distribution, which is used as the condition on $\m{X}$ to prove the sure screening property of FS by Wang
(2009). This and other results on comparisons of our assumptions with various conditions used in the literature will be discussed and reported elsewhere.

\begin{assumption}\label{as:nonsing} As $n\rightarrow\infty$, $(\delta_n|\beta_{\min}|^2)^{-1}=O(n^{\gamma_1}),\
\|\vg{\beta}\|(\delta_n|\beta_{\min}|^2)^{-1}=O(n^{\gamma_2}), \ M=O(n^{\gamma_3})$, and $\log(p)=O(n^{\gamma_4})$, where $\gamma_i\geqslant0,\ i=1,\ldots,4$,
$2\gamma_1+\gamma_3+\gamma_4<1$, and $2\gamma_2+2\gamma_3+\gamma_4<1$.\label{as:me}\end{assumption}

Theorem \ref{th:bbp} shows the ``better fitting, better screening" rule for variable screening, which means that, with probability tending to 1, a submodel that includes
$\s{A}_0$ yields smaller sum of squares than any submodel of the same size that does not.
\begin{theorem}\label{th:bbp}Under Assumption \ref{as:norm}, \ref{as:stand}, and \ref{as:me}, if $M\geqslant d$,
then as $n\rightarrow\infty$,$$\P\left(\max_{\s{A}\in \mathfrak{A}_0}\big\|\v{y}-\m{X}\hat{\vg{\beta}}^{\s{A}}\big\|^2<\min_{\s{A}\in
\mathfrak{A}_1}\big\|\v{y}-\m{X}\hat{\vg{\beta}}^{\s{A}}\big\|^2\right)=1-O\left(\exp(-C_1n^{\nu})\right),$$where $\nu=\min\{1-(2\gamma_1+\gamma_3+\gamma_4),\
1-(2\gamma_2+2\gamma_3+\gamma_4)\}$ and $C_1>0$ is a constant.\end{theorem}

For an $M$-subset $\s{T}$ of $\mathbb{Z}_p$, we call $\s{T}$ a superior subset if $\s{T}$ is better (in the sense of model fitting) than at least $|\mathfrak{A}_1|$
$M$-subsets. The ratio of all the superior subsets to all $M$-subsets is $|\mathfrak{A}_0|/|\mathbb{Z}_p|={p-d \choose M-d}/{p \choose M}={M \choose d}/{p \choose d}$,
which is increasing on $M\in(2d,n)$ for fixed $p$ and $d<n/2$. Theorem \ref{th:bbp} indicates that the set of all superior subsets is asymptotically identical to
$\mathfrak{A}_0$, which is stated as the following sure screening property of superior subsets.
\begin{corollary}\label{th:sc}Under the same conditions as in Theorem \ref{th:bbp}, for any superior subset $\s{T}$,
we have$$\P(\s{T}\supset\s{A}_0)=1-O\left(\exp(-C_1n^{\nu})\right)$$ as $n\rightarrow\infty$, where $\nu$ and $C_1$ are
the same as in Theorem \ref{th:bbp}.\end{corollary}

It is clear that the best subset from the solution to (\ref{bss}) is a superior subset. By Corollary
\ref{th:sc}, the best subset has the sure screening property.

Usually it is difficult to determine whether a given subset is a superior subset. However, from Theorem \ref{th:bbp} we can at least draw a conclusion that, for two
subsets with the same size, the better one (in the sense of model fitting) is more likely to be a superior subset, and thus is more likely to include the true submodel
asymptotically. This result is stated as the following corollary.

\begin{corollary}\label{th:bs}For two subsets $\s{T}_1$ and $\s{T}_2$ with $|\s{T}_1|=|\s{T}_2|=M$, suppose that
$\big\|\v{y}-\m{X}\hat{\vg{\beta}}^{\s{T}_1}\big\|^2\leqslant\big\|\v{y}-\m{X}\hat{\vg{\beta}}^{\s{T}_2}\big\|^2$. Then under the same conditions as in Theorem
\ref{th:bbp},
$$\liminf_{n\to\infty}\big[\P(\s{T}_1\supset\s{A}_0)-\P(\s{T}_2\supset\s{A}_0)\big]\geqslant0.$$\end{corollary}

From Corollary \ref{th:bs} we know that if $\s{T}_2$ has the sure screening property, then a better subset $\s{T}_1$ also has this property.

After screening $M$ variables by better subset regression, we can estimate the coefficients of the corresponding submodel by well-developed regression techniques for
situations where the variables are fewer than the observations. It is desirable to use a regularization method that can improve on least squares regression in terms of
variable selection and estimation accuracy. Such methods include the nonnegative garrote (Breiman, 1995), the lasso (Tibshirani, 1996), SCAD (Fan and Li, 2001), the
adaptive lasso (Zou 2006), and MCP (Zhang 2010). Xiong (2010) presented some advantages of the nonnegative garrote in interpretation and implementation. The ridge
regression-based nonnegative garrote method can have good performance even when the variables are highly correlated.

\section{Orthogonalizing subset screening}\label{sec:os}

\subsection{Orthogonalizing subset screening: an EM algorithm}\label{subsec:oem}

From the previous section, the superior subsets with good model fitting have the sure screening property. To obtain a superior subset, in this section we consider the
optimization problem (\ref{bss}) that yields the best subset. A new iterative algorithm, called orthogonalizing subset screening (OSS), is proposed for solving
(\ref{bss}). Since (\ref{bss}) is an N-P hard problem, our algorithm cannot guarantee achieving the best subset. Fortunately, with an appealing monotonicity property,
OSS improves the model fitting of an initial sparse estimator, and thus improves its asymptotic screening performance by Corollary \ref{th:bs}.

\begin{figure}
\centering \scalebox{0.4}[0.4]{\includegraphics{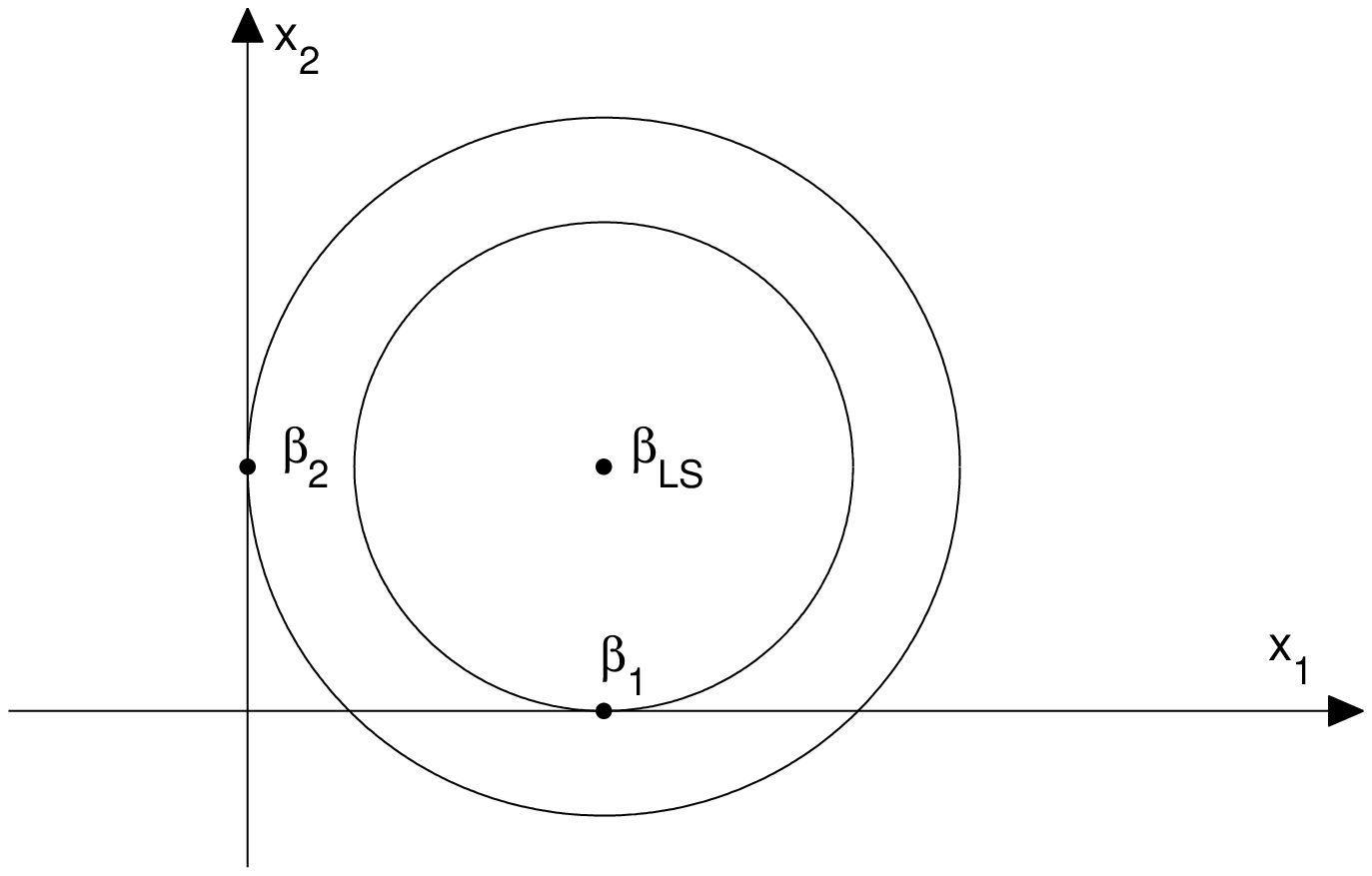}}\scalebox{0.4}[0.4]{\includegraphics{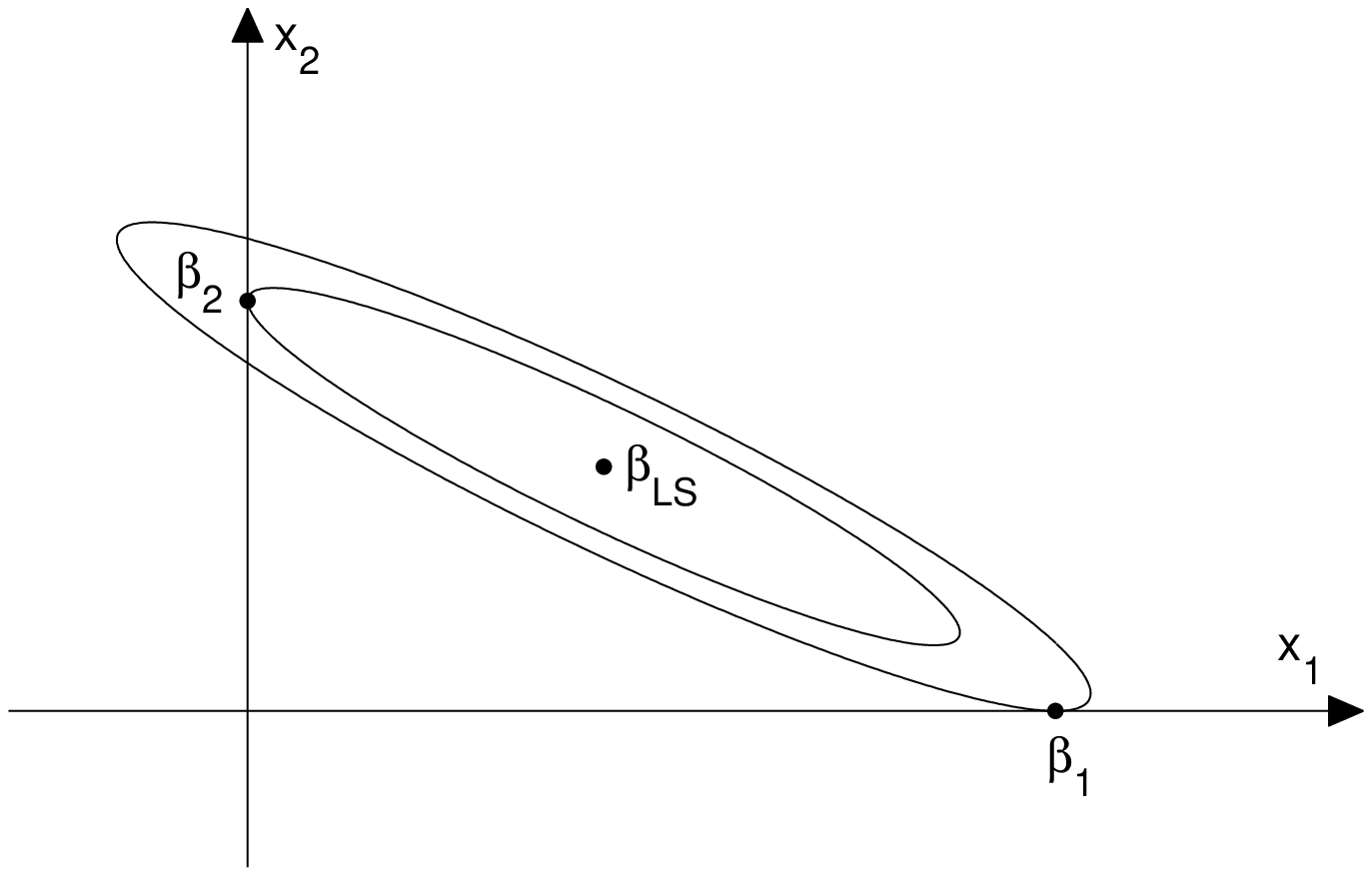}} \caption{\label{fig:comorth}The solution to (\ref{bss})
with $M=1$ in the case of two variables, where the circles and ellipses are contours of the objective function in (\ref{bss}), and $\vg{\beta}_{\mathrm{LS}}$,
$\vg{\beta}_1$ and $\vg{\beta}_2$ denote the least squares estimators under the full model and two submodels, respectively. On the left-hand side, the regression matrix
is orthogonal. The solution to (\ref{bss}) is $\vg{\beta}_1$, which corresponds to the larger component of $\vg{\beta}_{\mathrm{LS}}$ (see Theorem \ref{th:orth}). This
is not the case when the regression matrix is nonorthogonal. The right-hand side shows an example that the larger component of $\vg{\beta}_{\mathrm{LS}}$ does not
correspond to the solution to (\ref{bss}).}
\end{figure}

Define\begin{equation*}f(\vg{\beta})=\|\v{y}-\m{X}\vg{\beta}\|^2,\end{equation*}which is the objective function in (\ref{bss}). For a vector
$\v{x}=(x_1,\ldots,x_p)'\in\mathbb{R}^p$, let $\s{U}$ denote the set of the subscripts corresponding to the $M$ largest values of $|x_j|$'s. Define a map
$\v{z}=(z_1,\ldots,z_p)'=S_M(\v{x})$ to be $z_j=x_j$ for $j\in\s{U}$ and $z_j=0$ otherwise. If $S_M(\v{x})$ has multiple values, we take it to be an arbitrary fixed
value among them.

\begin{proposition}\label{th:orth} If $\m{X}'\m{X}=c\m{I}$, then $\vg{\beta}^*=S_M(\m{X}'\v{y})/c$ is a solution to (\ref{bss}).\end{proposition}

Figure \ref{fig:comorth} shows the difference between orthogonal and nonorthogonal cases for solving (\ref{bss}) when $p=2$ and $M=1$.

Proposition \ref{th:orth} inspires us to embed (\ref{bss}) into a problem with (column) orthogonal regression matrix. This idea, called active orthogonalization, was
proposed by Xiong, Dai and Qian (2011) for computing penalized least squares estimators. Here we apply it to (\ref{bss}). Take a number
$c\geqslant\lambda_{\max}(\m{X}'\m{X})$. Note that $c\m{I}-\m{X}'\m{X}\geqslant\m{0}$. Let $\m{\Delta}$ be a matrix satisfying
$\m{\Delta}'\m{\Delta}=c\m{I}-\m{X}'\m{X}$. Therefore
$$\m{X}_c=\left(\begin{array}{c}\m{X}\\ \m{\Delta}\end{array}\right)$$ is orthogonal. Consider the following linear model
\begin{equation}\m{y}_c=\m{X}_c\vg{\beta}+\vg{\varepsilon}_c,\label{lmc}\end{equation}where $\m{y}_c=(\m{y}',\m{y}_m')'$ is the complete response vector
including a missing part $\m{y}_m$. Based on the complete model in (\ref{lmc}), we can solve (\ref{bss}) by iteratively imputing $\m{y}_m$. Let $\vg{\beta}^{(0)}$
be an initial point. For $k=0,1,\ldots$, impute $\m{y}_m$ as $\m{y}_{imp}=\m{\Delta}\vg{\beta}^{(k)}$, let $\v{y}_{c,imp}=(\m{y}',\ \m{y}_{imp}')'$ and solve
\begin{equation}\min_{\vg{\beta}}\|\v{y}_{c,imp}-\m{X}_c\vg{\beta}\|^2\quad\text{subject to}\ \|\vg{\beta}\|_0\leqslant M.\label{bssc}\end{equation}
Since $\m{X}_c$ is orthogonal, the above problem has a closed-form solution by Proposition \ref{th:orth}. This leads to the following iteration formula
\begin{equation}\vg{\beta}^{(k+1)}=S_M\left(c^{-1}\m{X}'\v{y}+(\m{I}-c^{-1}\m{X}'\m{X})\vg{\beta}^{(k)}\right).\label{oem}\end{equation}
We call this algorithm orthogonalizing subset screening (OSS). In this paper, we always set $c$ in (\ref{oem}) to be $\lambda_{\max}(\m{X}'\m{X})$, which can be computed
by the power method (Wilkinson, 1965).

It can be seen that the OSS algorithm is an EM algorithm (Dempster, Laird and Rubin, 1977). Assume that the complete data $\m{y}_c=(\m{y}',\m{y}_m')'$ follows a normal
distribution $N(\m{X}_c\vg{\beta},\ \m{I})$. The likelihood function is $$L(\vg{\beta}\mid
\v{y})=(2\pi)^{-n/2}\exp\left(-\frac{1}{2}\|\v{y}-\m{X}\vg{\beta}\|^2\right).$$ Given $\vg{\beta}^{(k)}$, the E-step of the EM algorithm
is$$\E\left[\log\{L(\vg{\beta}|\v{y}_c)\}\mid
\v{y},\vg{\beta}^{(k)}\right]=-n\log(2\pi)/2-\|\v{y}-\m{X}\vg{\beta}\|^2/2-\|\m{\Delta}\vg{\beta}^{(k)}-\m{\Delta}\vg{\beta}\|^2/2.$$The M-step of the EM algorithm is to
minimize the above expectation subject to the constraint $\|\vg{\beta}\|_0\leqslant M$, which is equivalent to (\ref{bssc}).

Unlike FS that always tracks one path, different choices of initial points make OSS very flexible. We can even take a point that does not satisfy the constraint
$\|\vg{\beta}\|_0\leqslant M$ to be the initial point of the OSS algorithm. Since (\ref{bss}) often has many local minima, a multiple-initial-point scheme can be used in
the OSS algorithm to obtain a relatively good solution. For each initial point, we conduct OSS until it converges. The solution corresponding to the smallest value of
the objective function will be taken to be the final estimator.

\subsection{Monotonicity of OSS}\label{subsec:mono}

Write\begin{equation}\vg{\beta}^{(k+1)}=T_M(\vg{\beta}^{(k)}),\label{T}\end{equation} where the map $T_M$ is defined by (\ref{oem}). Like other EM algorithms, OSS has
the monotonicity property, which is stated below.
\begin{proposition}For any $\vg{\beta}\in\mathbb{R}^p$ with $\|\vg{\beta}\|_0\leqslant M$, $f(T_M(\vg{\beta}))\leqslant f(\vg{\beta})$.\label{th:mono}\end{proposition}

By Proposition \ref{th:mono}, the iterative map $T_M$ in OSS can improve on any sparse estimator in terms of model fitting. This also leads to an improvement of
asymptotic screening performance by the ``better fitting, better screening" rule. Specifically, let $\tilde{\vg{\beta}}=(\tilde{\beta}_1,\ldots,\tilde{\beta}_p)'$ be a
sparse estimator with $\|\tilde{\vg{\beta}}\|_0\leqslant M$. For any $\s{T}_0\subset\mathbb{Z}_p$ with $|\s{T}_0|=M$ and $\s{T}_0\supset\{j\in\mathbb{Z}_p:\
\tilde{\beta}_j\neq0\}$, the OSS sequence $\{T_M^{(k)}(\hat{\vg{\beta}}^{\s{T}_0})\}$ reduces the sum of squares step by step. When the iterative process stops by some
stopping rule in the $K$th iteration, we take $\hat{\vg{\beta}}=(\hat{\beta}_1,\ldots,\hat{\beta}_p)'=T_M^{(K)}(\hat{\vg{\beta}}^{\s{T}_0})$ to be a new estimator of
$\vg{\beta}$. After the improvement process, the final estimator $\hat{\vg{\beta}}=(\hat{\beta}_1,\ldots,\hat{\beta}_p)'$ fits the model better. Additionally, the
corresponding subset $\s{T}_{\mathrm{OSS}}=\{j\in\mathbb{Z}_p:\ \hat{\beta}_j\neq0\}$ has better asymptotic screening performance than the initial subset $\s{T}_0$ by
Corollary \ref{th:bs}.

Here we state some connections between OSS and other variable selection methods. When the initial point $\vg{\beta}^{(0)}$ in (\ref{oem}) is the zero vector, the
submodel selected by $\vg{\beta}^{(1)}$ is the same as that selected by Fan and Lv (2008)'s sure independence screening (SIS). Unlike SIS, OSS will go on seeking better
submodels after this iteration. When the $p$ least squares estimators under one-dimensional submodels are taken to be initial points, OSS looks similar to the
$L_2$Boosting algorithm (B\"{u}hlmann and Hothorn, 2007; Zhao and Yu, 2007). Both of them can produce better fits by combining these simple regression estimators. A
difference between them is that, $L_2$Boosting successively enters the variables, whereas OSS keeps the same number of variables after each iteration as we want.
Besides, in virtue of its monotonicity property, OSS can give a further improvement to the estimator from $L_2$Boosting by using the method in the previous paragraph.

\subsection{Convergence properties of OSS}\label{subsec:con}

This subsection focuses on convergence properties of OSS. By Proposition \ref{th:mono}, we can immediately obtain the monotonic convergence property of
$\{f(\vg{\beta}^{(k)})\}$.
\begin{proposition}Let $\{\vg{\beta}^{(k)}\}$ be a sequence generated by (\ref{oem}). For any $\vg{\beta}^{(0)}\in\mathbb{R}^p$, $\{f(\vg{\beta}^{(k)})\}_{k\geqslant1}$
converges monotonically to a limit as $k\rightarrow\infty$. \label{th:conv}\end{proposition}

Recall that the map $T_M$ in (\ref{T}) is not continuous. Although almost all OSS sequences converge in our numerical studies, counter examples exist in some special
cases. By Proposition \ref{th:conv}, we can stop an OSS iteration in (\ref{T}) when the sum of squares does not decrease numerically any more.

The general tools for proving the convergence of an EM algorithm (Zangwill, 1969; Wu, 1983) are not applicable to OSS because of the discontinuity of $T_M$. However, it
is possible to obtain good convergence properties of OSS under certain conditions. The following theorem shows that an OSS sequence can converge to the global solution
when the initial point lies in a neighborhood of the global solution. Recall that (\ref{bss}) is equivalent to a combinatorial optimization problem (\ref{bss2}). This
theorem makes OSS similar to an effective algorithm for a continuous nonconvex optimization problem.
\begin{theorem}Let $\{\vg{\beta}^{(k)}\}$ be a sequence generated by (\ref{oem}). Suppose that the problem (\ref{bss}) has a unique solution
denoted by $\vg{\beta}^*$ with $\|\vg{\beta}^*\|_0=M$. Then there exists a neighborhood $D\subset\mathbb{R}^p$ of $\vg{\beta}^*$ such that, for any $\vg{\beta}^{(0)}\in
D$, $\vg{\beta}^{(k)}\rightarrow\vg{\beta}^*$ as $k\rightarrow\infty$. \label{th:gconv}\end{theorem}

In practice, it is difficult to locate the neighborhood of $\vg{\beta}^*$ required in Theorem \ref{th:gconv}. However, this theorem still provides us a direction to
search $\vg{\beta}^*$. When $n$ is sufficiently large and the true $\vg{\beta}$ is sparse, we know that $\vg{\beta}^*$ is close to the true $\vg{\beta}$. Therefore, a
consistent estimator of $\vg{\beta}$, which is obtained by a computationally inexpensive method, can be used as the initial point in OSS to approach $\vg{\beta}^*$.
Using this way, we are more likely to obtain better subsets with good screening performance. For example, the lasso and SCAD are consistent under some regularity
conditions even when $p$ is much larger than $n$ (B\"{u}hlmann and van de Geer 2011; Fan and Lv 2011).

\subsection{Fast orthogonalizing subset screening}\label{subsec:foep}

Like other EM algorithm, a disadvantage of OSS is its sometimes very slow convergence. Here we provide a method to speed up the OSS algorithm. Note that the least
squares estimator yields the least sum of squares under a submodel. To avoid superfluous iterations of OSS in achieving the least squares estimator, we replace the
iteration formula (\ref{oem}) by
\begin{eqnarray}&&\vg{\phi}^{(k)}=S_M\left(c^{-1}\m{X}'\v{y}+(\m{I}-c^{-1}\m{X}'\m{X})\vg{\beta}^{(k)}\right),\nonumber\\&&\s{A}^{(k)} =\{j\in \mathbb{Z}_p:\
\phi_j^{(k)}\neq0\},\nonumber\\&&\vg{\beta}^{(k+1)}=\big(\m{X}_{\s{A}^{(k)}}'\m{X}_{\s{A}^{(k)}}\big)^{+}\m{X}_{\s{A}^{(k)}}'\v{y},\label{foem}\end{eqnarray} where
``$^+$" denotes the Moore-Penrose generalized inverse. We call this algorithm fast orthogonalizing subset screening (FOSS).

Denote the map from $\vg{\beta}^{(k)}$ to $\vg{\beta}^{(k+1)}$ in (\ref{foem}) by $T_M^F$. It follows from Proposition \ref{th:mono} and (\ref{foem}) that
$f(T_M^F(\vg{\beta}))\leqslant f(T_M(\vg{\beta}))\leqslant f(\vg{\beta})$ for any $\vg{\beta}\in\mathbb{R}^p$ with $\|\vg{\beta}\|_0\leqslant M$. This result not only
shows the monotonicity property of FOSS, but also indicates that FOSS can converge faster than OSS. Empirical results show that FOSS can obtain similar results to OSS
with far fewer iteration times.

\section{Simulations}\label{sec:sim}

\subsection{Deterministic design cases}\label{subsec:sd}

Supersaturated designs are commonly used in screening experiments for studying large-scale systems. In this simulation study, the design matrix $\m{X}$ in (\ref{lm}) is
taken as supersaturated designs, and the coefficients are given by $\beta_1=\cdots=\beta_5=1$ and $\beta_j=0$ for other $j$. The supersaturated designs are constructed
from the Kronecker tensor product of a small two-level supersaturated design with $n=12$ and $p=66$ in Wu (1993) and two $m\times m$ Hadamard matrices (Agaian 1985).
Here $m=2$ and $m=4$ are considered. Therefore we have two configurations of $n$ and $p$: $n=24,\ p=132$ and $n=48,\ p=264$.

\begin{table}
\caption{\label{tab:sd}Simulation results in Section \ref{subsec:sd} ($M=10$)} \centering\vspace{5mm}
\begin{tabular}{*{7}{lcccccc}}\hline\multirow{2}{*}{\em Method}&\quad\quad&\multicolumn{2}{c}{$n=24$}&\quad\quad&\multicolumn{2}{c}{$n=48$}
\\\cline{3-10}&&CR&AO&\quad\quad&CR&AO\\\hline
        LAR&&0.167&38.15&&0.306&69.62\\
   FOSS-LAR&&0.277&16.67&&0.795&36.15\\\hline
        SIS&&0.013&19.78&&0.827&37.52\\
   FOSS-SIS&&0.080&13.28&&0.947&30.96\\\hline
       ISIS&&0.028&10.54&&0.974&26.17\\
  FOSS-ISIS&&0.038&9.955&&0.974&25.44\\\hline
         FS&&0.192&3.115&&0.982&16.61\\
    FOSS-FS&&0.192&3.096&&0.982&16.57\\\hline
\end{tabular}
\end{table}

The following four methods are considered as basic methods for comparisons: Efron et al. (2004)'s least angle regression (LAR), Fan and Lv (2008)'s SIS and iterative SIS
(ISIS), and FS. Besides their popularity in variable screening, the reason why we choose them is that the number of variables selected by them can be exactly controlled
to be a specified number. Hence, we can compare them with our FOSS algorithm at the same size of submodels. Corresponding to the basic methods, four FOSS type algorithms
are used in our simulations, which are denoted by FOSS-LAR, FOSS-SIS, FOSS-ISIS, and FOSS-FS, respectively. After LAR, SIS, and ISIS select $M$-dimensional submodels,
FOSS-LAR, FOSS-SIS, and FOSS-ISIS respectively use the least squares estimators under the corresponding submodels as initial points in the FOSS iteration (\ref{foem}),
and derive new $M-$dimensional submodels. To obtain better local solutions to (\ref{bss}), we use the multiple-initial-point scheme in FOSS-FS. The initial points are
set as the least squares estimators under $L$-dimensional submodels selected by FS with $L$ from $M-[p/10]$ to $\min\{M+[p/10],\ n\}$, where $[\cdot]$ denotes the floor
function. In this subsection, $M$ is fixed as $10$.

We use 1000 repetitions in the simulations. There are two criteria to evaluate the eight methods: coverage rate (CR) and average objective values (AO), which denote the
percentage of times a method that includes the true submodel and average value of the sums of squares over the 1000 repetitions, respectively. The simulation results are
presented in Table \ref{tab:sd}. It can be seen that, FOSS cannot only reduce the sum of squares, but also yield local solutions to (\ref{bss}) with better, or at least
the same, screening performance. In particular, the local solutions around LAR and SIS derived by FOSS significantly improve their CRs, respectively. It is also
worthwhile noting that the results for $n=24$ seem inconsistent with the ``better fitting, better screening" rule (FOSS-LAR has the best screening performance but larger
AO than FS and FOSS-FS). This may be due to the small $n$. When $n=48$, we can see that the results follow this rule better.

\subsection{Random design cases}\label{subsec:rd}

In the simulation we use the following model\begin{equation}Y=\beta_0+\beta_1X_1+\cdots+\beta_pX_p+\varepsilon,\label{lr}\end{equation} where $X_1,\ldots,X_p$ are $p$
predictors and $\varepsilon\sim N(0,1)$ is noise that is independent of the predictors. The predictors $(X_1,\ldots,X_p)'$ is generated from a multivariate normal
distribution $N(\v{0},\m{\Sigma})$ whose covariance matrix $\m{\Sigma}=(\sigma_{ij})_{p\times p}$ has entries $\sigma_{ii}=1,\ i=1,\ldots,p$ and $\sigma_{ij}=\rho,\
i\neq j$, where $\rho=0,\ 0.5,$ and $0.9$ are considered. The coefficients are given by $\beta_1=\cdots=\beta_d=3$ and $\beta_j=0$ for other $j$. We use two
configurations of $n$ and $p$, $n=50,\ p=50$ and $n=200,\ p=500$, which represent small sample cases and large sample cases, respectively. For each model, we simulate
1000 data sets.

We compute the CRs and AOs of the same eight methods with $M=30$ as in Section \ref{subsec:sd} and the results are shown in Table \ref{tab:prspar}. It is clear to see
that, (i) FOSS can improve the four basic methods in terms of CR in most cases, especially when $\rho=0$. (ii) When $\rho$ is large, FOSS-LAR, FOSS-SIS, and FOSS-ISIS
cannot give significant improvement in model fitting since there are many local solutions to (\ref{bss}). In spite of this, each of the three FOSS algorithms has at
least the same CRs as the corresponding basic method. Unlike them, FOSS-FS reduces the sum of squares of FS much for all $\rho$'s because of the multiple-initial-point
scheme. (iii) The ``better fitting, better screening" rule holds, especially in the large sample cases. FS, with the smallest sum of squares, usually has the largest CRs
among the four basic methods. FOSS-FS often performs better than FS when $d=20$.

Combining the theoretical and empirical studies, we draw the following conclusions. When $d$ is relatively small compared to $M$, the number of the subsets that include
the true submodel $\s{A}_0$ is large. By Theorem \ref{th:bbp}, we do not need to find a subset with very good fitting, and FS is satisfactory for this case. When $d$ is
relatively large, we have to find much better subset, and FOSS-FS can be applied. Note that in practice $d$ is unknown. We prefer to use FOSS-FS since FOSS usually
converges very fast and can yield better fitting.

\begin{table}
\caption{\label{tab:prspar}Simulation results in Section \ref{subsec:rd} ($M=30$)} \centering\vspace{5mm}
\begin{tabular}{*{12}{lclccccccccc}}
\multicolumn{12}{c}{$n=50,\ p=50$}
\\\hline\multirow{2}{*}{$d$}&\quad&\multirow{2}{*}{\em Method}
&\quad\quad&\multicolumn{2}{c}{$\rho=0$}&\quad\quad&\multicolumn{2}{c}{$\rho=0.5$}&\quad\quad&\multicolumn{2}{c}{$\rho=0.9$}
\\\cline{5-12}&&&&CR&AO&\quad\quad&CR&AO&\quad\quad&CR&AO\\\hline
10           &&LAR&&0.173&235.4&&0.996&15.83&&0.970&16.54\\
        &&FOSS-LAR&&0.946&16.54&&0.996&15.31&&0.970&16.44\\
 \cline{3-12}&&SIS&&0.565&87.02&&0.250&131.9&&0.229&45.27\\
        &&FOSS-SIS&&0.991&10.69&&0.475&89.62&&0.244&44.38\\
\cline{3-12}&&ISIS&&0.971&20.28&&0.874&27.03&&0.781&20.21\\
       &&FOSS-ISIS&&0.998&9.979&&0.911&23.33&&0.785&20.05\\
  \cline{3-12}&&FS&&1    &6.221&&1    &6.179&&0.984&6.299\\
         &&FOSS-FS&&1    &5.047&&1    &5.006&&0.850&5.028\\\hline
20           &&LAR&&0    &760.1&&0.262&113.6&&0.131&38.78\\
        &&FOSS-LAR&&0.286&160.2&&0.292&106.7&&0.134&38.68\\
 \cline{3-12}&&SIS&&0.003&465.5&&0.001&551.2&&0    &139.6\\
        &&FOSS-SIS&&0.558&78.57&&0.005&471.0&&0    &137.1\\
\cline{3-12}&&ISIS&&0.051&234.7&&0.020&220.0&&0.007&60.93\\
       &&FOSS-ISIS&&0.607&62.27&&0.050&193.7&&0.008&60.34\\
  \cline{3-12}&&FS&&0.800&23.12&&0.804&17.53&&0.456&11.55\\
         &&FOSS-FS&&0.897&12.20&&0.904&10.25&&0.464&7.987\\\hline
\\\multicolumn{12}{c}{$n=200,\ p=500$}
\\\hline\multirow{2}{*}{$d$}&\quad&\multirow{2}{*}{\em Method}&
\quad\quad&\multicolumn{2}{c}{$\rho=0$}&\quad\quad&\multicolumn{2}{c}{$\rho=0.5$}&\quad\quad&\multicolumn{2}{c}{$\rho=0.9$}
\\\cline{5-12}&&&&CR&AO&\quad\quad&CR&AO&\quad\quad&CR&AO\\\hline
10           &&LAR&&0.926&281.9&&0.979&177.8&&0.863&189.4\\
        &&FOSS-LAR&&1    &121.9&&0.979&177.3&&0.866&188.9\\
 \cline{3-12}&&SIS&&0.932&258.4&&0.016&2341 &&0.006&736.3\\
        &&FOSS-SIS&&1    &121.5&&0.082&2172 &&0.008&732.1\\
\cline{3-12}&&ISIS&&1    &127.1&&0.978&182.3&&0.884&182.0\\
       &&FOSS-ISIS&&1    &113.2&&0.982&179.0&&0.886&181.6\\
  \cline{3-12}&&FS&&1    &86.64&&1    &86.93&&1    &88.23\\
         &&FOSS-FS&&1    &85.19&&1    &84.84&&1    &85.53\\\hline
20           &&LAR&&0.001&7106 &&0    &6923 &&0    &1760 \\
        &&FOSS-LAR&&0.996&157.4&&0    &6882 &&0    &1759 \\
 \cline{3-12}&&SIS&&0.009&4064 &&0    &1171 &&0    &2811 \\
        &&FOSS-SIS&&0.998&153.6&&0    &1160 &&0    &2809 \\
\cline{3-12}&&ISIS&&0.878&339.8&&0    &4857 &&0    &1244 \\
       &&FOSS-ISIS&&0.999&144.2&&0    &4838 &&0    &1243 \\
  \cline{3-12}&&FS&&1    &114.7&&0.994&133.2&&0.990&133.9\\
         &&FOSS-FS&&1    &113.8&&1    &114.1&&1    &115.1\\\hline
\end{tabular}
\end{table}

\section{A real data example}\label{sec:re}

We apply our methods to analyze some CT image data. The dataset used here for illustrating our methods is a part of the whole dataset in Frank and Asuncion (2010), and
is also available from the author. The dataset was retrieved from a set of 225 CT images from a person. Each CT slice is described by two histograms in polar space. The
first histogram has 240 components, describing the location of bone structures in the image. The second histogram has 144 components, describing the location of air
inclusions inside of the body. Both histograms are concatenated to form the final feature vector. The response variable is relative location of an image on the axial
axis, which was constructed by manually annotating up to 10 different distinct landmarks in each CT volume with known location. More detailed description of the dataset
can be found in Graf et al. (2011). Among those 225 images, 200 of them are set as the training sample and the remaining 25 of them are set to be the test sample.

\begin{table}\label{tab:real}
\caption{Results in Section \ref{sec:re}}\centering\vspace{5mm} \begin{tabular}{*{9}{lcccc}}\hline \em \multirow{2}{*}{Method}&\quad\quad&\em Test&&\em Sum
\\&&\em error&\quad\quad&\em of squares\\\hline
FS      &&0.626&&3.881\\
FOSS-FS &&0.472&&2.421\\\hline
\end{tabular}
\end{table}

We use linear regression to analyze the relationship between the feature vector and the response. Here the sample size $n=200$, which is much less than the number of
variables, $p=384$, in the feature vector. There are high correlations between the variables. We want to select a small part of variables to simplify the model and to
improve the prediction accuracy. FS and FOSS-FS with $M=20$ are applied here since they have showed us good performance in the simulation studies. After obtaining a
submodel with 20 variables by FS or FOSS-FS, we compute the least squares estimator under the submodel. The test errors and numbers of selected variables corresponding
to FS and FOSS-FS are shown in Table \ref{tab:real}. Since the variables in the feature vector are highly correlated, the two subsets selected by the procedures are
quite different. FOSS-FS performs better than FS in terms of test error.

\section{Discussion}\label{sec:dis}

This paper extends best subset regression, a classical variable selection technique, to better subset regression. For a screening purpose, we do not need to find the
best subset, and a ``better" one is enough in most cases. From the discussion in Section \ref{sec:bbp} and \ref{sec:os}, the word ``better" here has two-fold meaning. In
theory, an $M$-subset can asymptotically include the true submodel if and only if it is better than at least $|\mathfrak{A}_1|$ other $M$-subsets in terms of model
fitting. This theoretical result is called the ``better fitting, better screening" rule. In implementation, ``better" lies in the monotonicity property of OSS and FOSS.
By the ``better fitting, better screening" rule, for two subsets with the same size, the better one (in the sense of model fitting) is more likely to include the true
submodel asymptotically. Therefore, OSS and FOSS can improve asymptotic screening performance of any initial subset in virtue of their monotonicity property. Simulation
results in Section \ref{sec:sim} show that FOSS usually yields subsets having better screening performance than the initial estimators given by popular screening
methods.

By the ``better fitting, better screening" rule, when the number of important variables $d$ is as large as the desired size of selected subsets $M$, only the best subset
can asymptotically include all important variables. This indicates that the best subset is the least fallible for variable selection (screening). Hence, the optimization
problem (\ref{bss}) that yields the best subset deserves more research. In theory, OSS can achieve the global solution to (\ref{bss}) with a good initial point.

A commonly used algorithm for solving (\ref{bss}) is FS, which has been shown to have good screening performance by Wang (2009) and our simulations in Section
\ref{sec:sim}. Since FS has a relatively small sum of squares compared to other existing screening methods, we can use the ``better fitting, better screening" rule to
explain why FS performs quite well. In addition, FS can be used as a good basic procedure. Based on it, we expect to find more satisfactory methods. For recent studies
on FS and its modifications, we refer the reader to Zhang (2011). This paper also provides the FOSS-FS method that gives an improvement on FS. Another advantage of FS is
that there are simple methods to choose $M$ when using it. Wang (2009) presents a BIC criterion in FS, and we can also use the $M$ selected from this criterion in
FOSS-FS. However, the choice of $M$ in better subset regression needs more investigation in the follow-up research.

Compared to best subset regression, better subset regression brings us a rule rather than a specific method. We are interested in whether
the ``better fitting, better screening" rule holds for more general cases and believe that this is a valuable topic in the future.

\section*{Acknowledgements}

This work is supported by the National Natural Science
Foundation of China (Grant No. 11271355). The author is also grateful to the support of Key Laboratory of Systems and Control, Chinese Academy of Sciences.

\section*{Appendix}

\begin{lemma}\label{lemma2}Let $\chi_n^2$ be a chi-square random variable with degrees of freedom $n$. We have \begin{equation}
\P\left(\frac{\chi_n^2}{n}\geqslant z\right)\leqslant\exp\left(-\frac{n(z-1)^2}{4z}\right)\quad\text{for}\quad z>1\label{chi1}\end{equation} and
\begin{equation}\P\left(\frac{\chi_n^2}{n}\leqslant z\right)\leqslant\exp\left(-\frac{n(1-z)^2}{4(2-z)}\right)\quad\text{for}\quad
z<1.\label{chi2}\end{equation}\end{lemma} The lemma can be proved by Bernstein's inequality (Uspensky, 1937), and its proof is omitted here.

\vspace{4mm}\noindent\emph{Proof of Theorem \ref{th:bbp}.} We have
\begin{eqnarray}&&\P\left(\max_{\s{A}\in \mathfrak{A}_0}\big\|\v{y}-\m{X}\hat{\vg{\beta}}^{\s{A}}\big\|^2<\min_{\s{A}\in
\mathfrak{A}_1}\big\|\v{y}-\m{X}\hat{\vg{\beta}}^{\s{A}}\big\|^2\right)\nonumber
\\&\geqslant&1-\sum_{\s{T}\in \mathfrak{A}_0}\sum_{\s{A}\in \mathfrak{A}_1}\P\left(\big\|\v{y}-\m{X}\hat{\vg{\beta}}^{\s{T}}\big\|^2
\geqslant\big\|\v{y}-\m{X}\hat{\vg{\beta}}^{\s{A}}\big\|^2\right).\nonumber
\\&\geqslant&1-\sum_{\s{T}\in \mathfrak{A}_0}\sum_{\s{A}\in \mathfrak{A}_1}\left[\P\left(\frac{\big\|\v{y}-\m{X}\hat{\vg{\beta}}^{\s{T}}\big\|^2}{(n-r_{\s{T}})\sigma^2}
\geqslant1+2\eta\right)+\P\left(\frac{\big\|\v{y}-\m{X}\hat{\vg{\beta}}^{\s{A}}\big\|^2}{(n-r_{\s{T}})\sigma^2}\leqslant1+2\eta\right)\right],
\label{key}\end{eqnarray}where $r_{\s{T}}$ is the rank of $\m{X}_{\s{T}}$ and $\eta=\delta_n|\beta_{\min}|^2/(4\sigma^2)$.

Note that $\big\|\v{y}-\m{X}\hat{\vg{\beta}}^{\s{T}}\big\|^2/\sigma^2\sim\chi_{n-r_{\s{T}}}^2$. By (\ref{chi1}),
\begin{equation}\P\left(\frac{\big\|\v{y}-\m{X}\hat{\vg{\beta}}^{\s{T}}\big\|^2}{(n-r_{\s{T}})\sigma^2}
\geqslant1+2\eta\right)\leqslant\exp\left(-\frac{\eta^2(n-r_{\s{T}})}{1+2\eta}\right)=O\left(\exp\left[-C_2(\delta_n|\beta_{\min}|^2)^2n\right]\right),
\label{firste}\end{equation} where $C_2>0$ is a constant.

Next we consider $\big\|\v{y}-\m{X}\hat{\vg{\beta}}^{\s{A}}\big\|^2$, which can be written as
\begin{eqnarray*}\big\|\v{y}-\m{X}\hat{\vg{\beta}}^{\s{A}}\big\|^2&=&\vg{\varepsilon}'\m{H}_{\s{A}}\vg{\varepsilon}
+2\vg{\beta}_{\s{A}_0}'\m{X}_{\s{A}_0}'\m{H}_{\s{A}}\vg{\varepsilon} +\vg{\beta}_{\s{A}_0}'\m{X}_{\s{A}_0}'\m{H}_{\s{A}}\m{X}_{\s{A}_0}\vg{\beta}_{\s{A}_0}
\\&=&\vg{\varepsilon}'\m{H}_{\s{A}}\vg{\varepsilon}
+2\vg{\beta}_{\s{A}_0}'\m{X}_{\s{A}_0}'\m{H}_{\s{A}}\vg{\varepsilon}+\vg{\beta}_{\s{A}_0\setminus\s{A}}'\m{X}_{\s{A}_0\setminus\s{A}}'
\m{H}_{\s{A}}\m{X}_{\s{A}_0\setminus\s{A}}\vg{\beta}_{\s{A}_0\setminus\s{A}}
\\&\geqslant&\vg{\varepsilon}'\m{H}_{\s{A}}\vg{\varepsilon}
+2\vg{\beta}_{\s{A}_0}'\m{X}_{\s{A}_0}'\m{H}_{\s{A}}\vg{\varepsilon}+n\delta_n|\beta_{\min}|^2.\end{eqnarray*}Note that
$\vg{\varepsilon}'\m{H}_{\s{A}}\vg{\varepsilon}/\sigma^2\sim\chi_{n-r_{\s{A}}}^2$ and $\vg{\beta}_{\s{A}_0}'\m{X}_{\s{A}_0}'\m{H}_{\s{A}}\vg{\varepsilon}\sim N(0,v^2)$,
where $r_{\s{A}}$ is the rank of $\m{X}_{\s{A}}$ and $v^2=\sigma^2\vg{\beta}_{\s{A}_0}'\m{X}_{\s{A}_0}' \m{H}_{\s{A}}\m{X}_{\s{A}_0}\vg{\beta}_{\s{A}_0}$. By Assumption
\ref{as:stand}, \begin{equation}v^2\leqslant \sigma^2\lambda_{\max}(\m{H}_{\s{A}})\lambda_{\max}(\m{X}_{\s{A}_0}'\m{X}_{\s{A}_0})\|\vg{\beta}\|^2 \leqslant
\sigma^2{\mathrm{tr}}(\m{X}_{\s{A}_0}'\m{X}_{\s{A}_0})\|\vg{\beta}\|^2\leqslant nCM\sigma^2\|\vg{\beta}\|^2.\label{v2}\end{equation} We have
\begin{eqnarray}&&\P\left(\frac{\big\|\v{y}-\m{X}\hat{\vg{\beta}}^{\s{A}}\big\|^2}{(n-r_{\s{T}})n\sigma^2}
\leqslant1+2\eta\right)\nonumber
\\&\leqslant&\P\left(\frac{\vg{\varepsilon}'\m{H}_{\s{A}}\vg{\varepsilon}}{n\sigma^2}
+\frac{2\vg{\beta}_{\s{A}_0}'\m{X}_{\s{A}_0}'\m{H}_{\s{A}}\vg{\varepsilon}}{n\sigma^2}\leqslant 1-2\eta\right)\nonumber
\\&\leqslant&\P\left(\frac{\vg{\varepsilon}'\m{H}_{\s{A}}\vg{\varepsilon}}{n\sigma^2}\leqslant 1-\eta\right)
+\P\left(\frac{2\vg{\beta}_{\s{A}_0}'\m{X}_{\s{A}_0}'\m{H}_{\s{A}}\vg{\varepsilon}}{n\sigma^2}\leqslant -\eta\right).\label{pA}\end{eqnarray} For sufficiently large $n$,
by (\ref{chi2}), \begin{eqnarray}&&\P\left(\frac{\vg{\varepsilon}'\m{H}_{\s{A}}\vg{\varepsilon}}{n\sigma^2}\leqslant 1-\eta\right)\leqslant
\P\left(\frac{\vg{\varepsilon}'\m{H}_{\s{A}}\vg{\varepsilon}}{(n-r_{\s{A}})\sigma^2}\leqslant 1-\eta/2\right)
\leqslant\exp\left(-\frac{\eta^2n}{16+8\eta}\right)\nonumber\\&&=O\left(\exp\left[-C_3(\delta_n|\beta_{\min}|^2)^2n\right]\right),\label{t1}\end{eqnarray} where $C_3>0$
is a constant. Denote the distribution function of the standard normal distribution by $\Phi$. Since $1-\Phi(x)<\exp(-x^2/2)/x$ for any $x>0$, by (\ref{v2}),
\begin{eqnarray}&&\P\left(\frac{2\vg{\beta}_{\s{A}_0}'\m{X}_{\s{A}_0}'\m{H}_{\s{A}}\vg{\varepsilon}}{n\sigma^2}\leqslant -\eta\right)
=1-\Phi\left(\frac{n\sigma^2\eta}{2v}\right)\leqslant\frac{2v}{n\sigma^2\eta}\exp\left(-\frac{n^2\sigma^4\eta^2}{8v^2}\right)\nonumber
\\&& =O\left(\frac{\sqrt{M}\|\vg{\beta}\|}{\sqrt{n}\delta_n|\beta_{\min}|^2}\exp\left(-\frac{C_4n\delta_n^2|\beta_{\min}|^4}{M\|\vg{\beta}\|^2}\right)\right),
\label{t2}\end{eqnarray}where $C_4>0$ is constant.

Note that $|\mathfrak{A}_0|\cdot|\mathfrak{A}_1|<p^{2M}$. Combining (\ref{key}), (\ref{firste}), (\ref{pA}), (\ref{t1}), and (\ref{t2}), we have
\begin{eqnarray*}&&\P\left(\max_{\s{A}\in \mathfrak{A}_0}\big\|\v{y}-\m{X}\hat{\vg{\beta}}^{\s{A}}\big\|^2<\min_{\s{A}\in
\mathfrak{A}_1}\big\|\v{y}-\m{X}\hat{\vg{\beta}}^{\s{A}}\big\|^2\right)
\\&>&1-p^{2M}\left[O\left(\exp\left[-C_5(\delta_n|\beta_{\min}|^2)^2n\right]\right)
+O\left(\frac{\sqrt{M}\|\vg{\beta}\|}{\sqrt{n}\delta_n|\beta_{\min}|^2}\exp\left(-\frac{C_4n\delta_n^2|\beta_{\min}|^4}{M\|\vg{\beta}\|^2}
\right)\right)\right],\end{eqnarray*}where $C_5=\min\{C_2,C_3\}$. By Assumption \ref{as:nonsing}, we complete the proof. $\quad\quad\quad\square$

\vspace{4mm}\noindent\emph{Proof of Corollary \ref{th:sc}.} We have $$\P(\s{T}\supset\s{A}_0)\geqslant\P\left(\max_{\s{A}\in
\mathfrak{A}_0}\big\|\v{y}-\m{X}\hat{\vg{\beta}}^{\s{A}}\big\|^2<\min_{\s{A}\in
\mathfrak{A}_1}\big\|\v{y}-\m{X}\hat{\vg{\beta}}^{\s{A}}\big\|^2\right).$$By Theorem \ref{th:bbp}, we complete the proof. $\quad\quad\quad\square$

\vspace{4mm}\noindent\emph{Proof of Corollary \ref{th:bs}.} Denote the set of all superior subsets by $\mathfrak{S}$. By Theorem \ref{th:bbp},
$\P(\mathfrak{S}=\mathfrak{A}_0)\to1$. It is not hard to show that
$$\P(\s{T}_1\in\mathfrak{A}_0)-\P(\s{T}_2\in\mathfrak{A}_0)=\P(\s{T}_1\in\mathfrak{S})-\P(\s{T}_2\in\mathfrak{S})+o(1).$$We complete the proof by noting that
$\P(\s{T}_1\in\mathfrak{S})\geqslant\P(\s{T}_2\in\mathfrak{S})$. $\quad\quad\quad\square$

\vspace{4mm}\noindent\emph{Proof of Proposition \ref{th:orth}.} Let $\hat{\vg{\beta}}=(\hat{\beta}_1,\ldots,\hat{\beta}_p)'=\m{X}'\v{y}/c$ be the least squares estimator
under $\m{X}'\m{X}=c\m{I}$. Note that $f(\vg{\beta})=c\|\vg{\beta}-\hat{\vg{\beta}}\|^2+\|\v{y}\|^2-\|\hat{\vg{\beta}}\|^2$. We only need to consider
$g(\vg{\beta})=\|\vg{\beta}-\hat{\vg{\beta}}\|^2$. For any $\vg{\beta}$ with $\|\vg{\beta}\|_0\leqslant M$, we have
\begin{eqnarray*}g(\vg{\beta})\geqslant\sum_{\hat{\beta}_j=0}\hat{\beta}_j^2\geqslant\sum_{\beta_j^*=0}\hat{\beta}_j^2=g(\vg{\beta}^*).\end{eqnarray*} This completes
the proof. $\quad\quad\quad\square$

\vspace{4mm}\noindent\emph{Proof of Proposition \ref{th:mono}.} Note that
$$T_M(\vg{\beta})=\arg\min_{\vg{\phi}}\big\{f(\vg{\phi})+\|\m{\Delta\vg{\phi}}-\m{\Delta}\vg{\beta}\|^2:\ \|\vg{\phi}\|_0\leqslant M\big\}.$$We
have\begin{eqnarray*}f(T_M(\vg{\beta}))\leqslant f(\vg{\beta})+\|\m{\Delta\vg{\beta}}-\m{\Delta}\vg{\beta}\|^2=f(\vg{\beta}),\end{eqnarray*}which completes the proof.
$\quad\quad\quad\square$

\vspace{4mm}\begin{lemma}If the problem (\ref{bss}) has a unique solution denoted by $\vg{\beta}^*$, then $\vg{\beta}^*$ is a fixed point of $T_M$, i.e.,
$\vg{\beta}^*=T_M(\vg{\beta}^*)$. \label{lemma:fp}\end{lemma}

\noindent\emph{Proof of Lemma \ref{lemma:fp}.} By Proposition \ref{th:mono}, $f(T_M(\vg{\beta}^*))\leqslant f(\vg{\beta}^*)$. Since the minimum is unique, we have
$\vg{\beta}^*=T_M(\vg{\beta}^*)$. $\quad\quad\quad\square$

\vspace{4mm}\noindent\emph{Proof of Theorem \ref{th:gconv}.} Without loss of generality, let $\s{A}^*=\{j\in\mathbb{Z}_p:\ \beta_j^*\neq0\}=\{1,\ldots,M\}$. Denote
$\s{B}^*=\mathbb{Z}_p\setminus\s{A}^*$. Define a function $u$ on $\mathbb{R}^M$ to be $u(x_1,\ldots,x_M)=\min\{|x_j|:\ j=1,\ldots,M\}$. By Lemma \ref{lemma:fp},
\begin{equation}\vg{\beta}^*=\left(\begin{array}{c}\vg{\beta}_{\s{A}^*}^*
\\\v{0}\end{array}\right)=T_M(\vg{\beta}^*)=S_M\left(\left(\begin{array}{c}c^{-1}\m{X}_{\s{A}^*}'\v{y}+(\m{I}-c^{-1}\m{X}_{\s{A}^*}'
\m{X}_{\s{A}^*})\vg{\beta}_{\s{A}^*}^*
\\c^{-1}\m{X}_{\s{B}^*}'\v{y}-c^{-1}\m{X}_{\s{B}^*}'\m{X}_{\s{A}^*}\vg{\beta}_{\s{A}^*}^*\end{array}\right)\right).\label{bstar}\end{equation}Since
$\vg{\beta}_{\s{B}^*}^*$ is the unique solution, (\ref{bstar}) implies
\begin{equation} u\big(c^{-1}\m{X}_{\s{A}^*}'\v{y}+(\m{I}-c^{-1}\m{X}_{\s{A}^*}'\m{X}_{\s{A}^*})\vg{\beta}_{\s{A}^*}^*\big)
>\big\|c^{-1}\m{X}_{\s{B}^*}'\v{y}-c^{-1}\m{X}_{\s{B}^*}'\m{X}_{\s{A}^*}\vg{\beta}_{\s{A}^*}^*\big\|_\infty,\label{bstarp}\end{equation}
where $\|\cdot\|_\infty$ denotes the $\ell_\infty$ norm.

Consider the set \begin{eqnarray*}&&E=\Big\{\vg{\beta}\in\mathbb{R}^p:\ u\big(c^{-1}\m{X}_{\s{A}^*}'\v{y}+(\m{I}
-c^{-1}\m{X}_{\s{A}^*}'\m{X}_{\s{A}^*})\vg{\beta}_{\s{A}^*}-c^{-1}\m{X}_{\s{A}^*}'\m{X}_{\s{B}^*}
\vg{\beta}_{\s{B}^*}\big)
\\&&\quad\quad\quad>\big\|c^{-1}\m{X}_{\s{B}^*}'\v{y}-c^{-1}\m{X}_{\s{B}^*}'\m{X}_{\s{A}^*}\vg{\beta}_{\s{A}^*}+(\m{I}-c^{-1}\m{X}_{\s{B}^*}'
\m{X}_{\s{B}^*})\vg{\beta}_{\s{B}^*}
\big\|_\infty\Big\}.
\end{eqnarray*}We have\begin{equation}T_M(\vg{\beta})
=\left(\begin{array}{c}c^{-1}\m{X}_{\s{A}^*}'\v{y}+(\m{I}-c^{-1}\m{X}_{\s{A}^*}'\m{X}_{\s{A}^*})\vg{\beta}_{\s{A}^*}-c^{-1}\m{X}_{\s{A}^*}'
\m{X}_{\s{B}^*}\vg{\beta}_{\s{B}^*}
\\\v{0}\end{array}\right)\quad\text{for}\ \vg{\beta}\in E.\label{E}\end{equation}By (\ref{bstarp}), $\vg{\beta}^*\in E$. Thus, there exists
$\delta>0$ such that the closed ball $\{\vg{\beta}:\ \|\vg{\beta}-\vg{\beta}^*\|\leqslant\delta\}\subset E$. Denote
$\nu=\max\big\{\lambda_{\max}(\m{I}-c^{-1}\m{X}_{\s{A}^*}'\m{X}_{\s{A}^*}), \
c^{-1}[\lambda_{\max}(\m{X}_{\s{B}^*}'\m{X}_{\s{A}^*}\m{X}_{\s{A}^*}'\m{X}_{\s{B}^*})]^{1/2}\big\}$ and $\tau=\min\{\delta,\ \delta/(\sqrt{2}\nu)\}$. Note that $\tau>0$.
For any $\vg{\beta}^{(0)}\in D=\{\vg{\beta}:\ \|\vg{\beta}-\vg{\beta}^*\|\leqslant\tau\}$, by (\ref{E}),\begin{eqnarray*}&&\|\vg{\beta}^{(1)}-\vg{\beta}^*\|
\\&=&\big\|(\m{I}-c^{-1}\m{X}_{\s{A}^*}'\m{X}_{\s{A}^*})(\vg{\beta}_{\s{A}^*}^{(0)}-\vg{\beta}_{\s{A}^*})-c^{-1}\m{X}_{\s{A}^*}'
\m{X}_{\s{B}^*}\vg{\beta}^{(0)}_{\s{B}^*}\big\|
\\&\leqslant&\nu\,\big(\big\|\vg{\beta}_{\s{A}^*}^{(0)}-\vg{\beta}_{\s{A}^*}^*\big\|+\big\|\vg{\beta}_{\s{B}^*}^{(0)}\big\|\big)
\\&\leqslant&\sqrt{2}\,\nu\,\big\|\vg{\beta}^{(0)}-\vg{\beta}^*\big\|
\\&\leqslant&\delta.\end{eqnarray*}Therefore, $\vg{\beta}^{(1)}\in E$. Consider $\vg{\beta}^{(2)}$, we have
\begin{eqnarray}&&\|\vg{\beta}^{(2)}-\vg{\beta}^*\|\nonumber
\\&=&\big\|(\m{I}-c^{-1}\m{X}_{\s{A}^*}'\m{X}_{\s{A}^*})(\vg{\beta}_{\s{A}^*}^{(1)}-\vg{\beta}_{\s{A}^*}^*)\big\|\nonumber
\\&\leqslant&\lambda_{\max}(\m{I}-c^{-1}\m{X}_{\s{A}^*}'\m{X}_{\s{A}^*})\big\|\vg{\beta}^{(1)}-\vg{\beta}^*\big\|.\label{b2}\end{eqnarray}
Since $\lambda_{\max}(\m{I}-c^{-1}\m{X}_{\s{A}^*}'\m{X}_{\s{A}^*})<1$, (\ref{b2}) implies $\vg{\beta}^{(2)}\in E$. By induction, we can prove that for
$k\geqslant2$, $\vg{\beta}^{(k)}\in E$ and $$\|\vg{\beta}^{(k)}-\vg{\beta}^*\|\leqslant\lambda_{\max}(\m{I}-c^{-1}\m{X}_{\s{A}^*}'\m{X}_{\s{A}^*})
\big\|\vg{\beta}^{(k-1)}-\vg{\beta}^*\big\|,$$which implies $\vg{\beta}^{(k)}\rightarrow\vg{\beta}^*$ as $k\rightarrow\infty$.
$\quad\quad\quad\square$

\vspace{1cm}
\noindent{\Large\bf References}

{\footnotesize \begin{description}

\item{}
Agaian, S. S. (1985) \textit{Hadamard matrices and their applications}. Berlin: Springer.



\item{}
Beale, E. M. L., Kendall, M. G. and Mann, D. W. (1967)
``The Discarding of Variables in Multivariate Analysis,"
\textit{Biometrika}, {\bf 54}, 357--366.



\item{}
Breiman, L. (1995)
``Better Subset Regression Using the Nonnegative Garrote,"
\textit{Technometrics}, {\bf 37}, 373--384.

\item{}
B\"{u}hlmann, P. and Hothorn, T. (2007)
``Boosting Algorithms: Regularization, Prediction and Model Fitting,"
\textit{Statistical Science}, {\bf 22}, 477--505.

\item{}
B\"{u}hlmann, P. and van de Geer, S. (2011) \textit{Statistics for High-Dimensional Data: Methods, Theory and Applications}, New York: Springer.

\item{}
Dempster, A. P., Laird, N. M. and Rubin, D. B. (1977)
``Maximum Likelihood from Incomplete Data via the EM Algorithm,"
\textit{Journal of the Royal Statistical Society, Ser. B}, {\bf 39}, 1--38.

\item{}
Efron, B., Hastie, T., Johnstone, I. and Tibshirani, R. (2004)
``Least Angle Regression,"
\textit{The Annals of Statistics}, \textbf{32}, 407--451.

\item{}
Fan, J. and Li, R. (2001)
``Variable Selection via Nonconcave Penalized Likelihood and Its Oracle Properties,"
\textit{Journal of the American Statistical Association}, \textbf{96}, 1348--1360.

\item{}
Fan, J. and Lv, J. (2008)
``Sure Independence Screening for Ultrahigh Dimensional Feature Space (with discussion),"
\textit{Journal of the Royal Statistical Society, Ser. B}, \textbf{70}, 849--911.

\item{}
Fan, J. and Lv, J. (2011). ``Properties of Non-concave Penalized Likelihood with NP-dimensionality," \textit{Information Theory, IEEE Transactions}, 57, 5467--5484.

\item{}
Fan, J., Samworth, R. and Wu, Y. (2009) ``Ultrahigh dimensional variable selection: beyond the linear model" \textit{Journal of Machine Learning Research}, \textbf{10},
2013--2038.

\item{}
Fan, J. and Song, R. (2010) ``Sure independence screening in generalized linear models with NP-dimensionality" \textit{The Annals of Statistics}, \textbf{38},
3567--3604.


\item{}
Frank, A. and Asuncion, A. (2010)
\textit{UCI Machine Learning Repository}
Irvine, CA: University of California, School of Information and Computer Science.\\ http://archive.ics.uci.edu/ml.

\item{}
Furnival, G. and Wilson, R. (1974)
``Regressions by Leaps and Bounds,"
\textit{Technometrics}, \textbf{16}, 499--511.

\item{}
Gatu, C. and Kontoghiorghes, E. J. (2006)
``Branch-and-Bound Algorithms for Computing the Best-Subset Regression Models,"
\textit{Journal of Computational and Graphical Statistics}, \textbf{15}, 139--156.


\item{}
Graf, F., Kriegel, H.-P., Schubert, M., P\"{o}elsterl, S. and Cavallaro, A. (2011) ``2D Image Registration in CT Images using Radial Image Descriptors," \textit{Lecture
Notes in Computer Science.}, \textbf{6892}, 607--614.

\item{}
Hall, P. and Miller, H. (2009) ``Using generalized correlation to effect variable selection in very high dimensional problems," \textit{Journal of Computational and
Graphical Statistics}, \textbf{18}, 533--550.


\item{}
Hocking, R. R. and Leslie, R. N. (1967)
``Selection of the Best Subset in Regression Analysis,"
\textit{Technometrics}, \textbf{9}, 531--540.





\item{}
LaMotte, L. R. and Hocking, R. R. (1970)
``Computational Efficiency in the Selection of Regression Variables,"
\textit{Technometrics}, \textbf{12}, 83--93.

\item{}
Li, G. Peng, H. Zhang, J. and Zhu, L. (2012) ``Robust rank correlation based screening" \textit{The Annals of Statistics}, \textbf{40}, 1846--1877.

\item{}
Lin, D. K. J. (1993), ``A New Class of Supersaturated Designs," \textit{Technometrics}, 35, 28--31.


\item{}
Miller, A. (2002)
\textit{{Subset Selection in Regression, 2nd Edition}}. Chapman \& Hall/CRC.

\item{}
Narendra, P. M. and Fukunaga, K. (1977)
``A Branch and Bound Algorithm for Feature Subset Selection,"
\textit{IEEE Transactions on Computers}, \textbf{26}, 917--922.



\item{}
Tibshirani, R. (1996)
``Regression Shrinkage and Selection via the Lasso,"
\textit{Journal of the Royal Statistical Society, Ser. B}, \textbf{58}, 267--288.


\item{}
Uspensky, J. V. (1937)
\textit{Introduction to Mathematical Probability}. McGraw-Hill Book Company,

\item{}
Wang, H. (2009)
``Forward Regression for Ultra-High Dimensional Variable Screening,"
\textit{Journal of the American Statistical Association}, \textbf{104}, 1512--1524.

\item{}
Wilkinson, J. H. (1965)
\textit{The Algebraic Eigenvalue Problem}. New York: Oxford University Press.

\item{}
Wu, C. F. J. (1983)
``On the Convergence Properties of the EM Algorithm,"
\textit{The Annals of Statistics}, \textbf{11}, 95--103.

\item{}
Wu, C. F. J. (1993), ``Construction of Supersaturated Designs through Partially Aliased Interactions," \textit{Biometrika}, 80, 661--669.

\item{}
Xiong, S. (2010)
``Some Notes on the Nonnegative Garrote,"
\textit{Technometrics}, \textbf{52}, 349-361.

\item{}
Xiong, S., Dai, B. and Qian, P. Z. G. (2011)
``Orthogonalizing Penalized Regression,"
\textit{Technical Report}, available at http://arxiv.org/PS$\_$cache/arxiv/pdf/1108/1108.0185v1.pdf.



\item{}
Zangwill, W. I. (1969)
\textit{Nonlinear Programming: A Unified Approach}. Englewood Cliffs, New Jersey: Prentice Hall.

\item{}
Zhang. C-H. (2010), ``Nearly Unbiased Variable Selection under Minimax Concave Penalty," \textit{The Annals of Statistics}, 38, 894--942.

\item{}
Zhang, T. (2011)
``Adaptive Forward-Backward Greedy Algorithm for Learning Sparse Representations,"
\textit{IEEE Trans. Inform. Theory}, \textbf{57}, 4689--4708.


\item{}
Zhao, P. and Yu, B. (2007)
``Stagewise Lasso,"
\textit{Journal of Machine Learning Research}, \textbf{8}, 2701--2726.


\item{}
Zou, H. (2006)
``The Adaptive Lasso and Its Oracle Properties,"
\textit{Journal of the American Statistical Association}, \textbf{101}, 1418--1429.

\end{description}}

\end{document}